\documentstyle[preprint,psfig,aps]{revtex} 
\def\be{\begin{equation}}
\def\ee{\end{equation}} \def\e{\epsilon} 
\def\bea{\begin{eqnarray}}
\def\eea{\end{eqnarray}}
\begin{document} 
\title{Cluster
ionization via two-plasmon excitation} 
\author{G.F.  Bertsch$^{(a)}$,  N. Van Giai$^{(b)}$ 
and N. Vinh Mau$^{(b)}$} 
\address{$^{(a)}$Dept. of Physics and Institute for Nuclear Theory,
Box 351560\\
University of Washington
Seattle, WA 98195\\}
\address{$^{(b))}$Groupe de Physique Th\'eorique\\
Institut de Physique Nucl\'eaire, 91406-Orsay Cedex, France}

\maketitle
\begin{abstract} We calculate the two-photon ionization of clusters
for photon energies near the surface plasmon resonance.  The results
are expressed in terms of the ionization rate of a double plasmon
excitation,
which is calculated perturbatively.  For the conditions of the experiment 
by Schlipper et al., \cite{sc98}, we find an ionization rate of the order
of 0.05-0.10 fs$^{-1}$.  This rate is used to determine the ionization
probability in an external field in terms of the number of photons absorbed
and the duration of the field.  The probability also depends on the damping
rate of the surface plasmon.  Agreement with experiment can only be
achieved if the plasmon damping is considerably smaller than its observed
width in the room-temperature single-photon absorption spectrum.
\end{abstract} 
\section{Introduction}
  The electromagnetic response of alkali metal clusters shows
a very strong surface plasmon resonance\cite{de93}, but the interactions
of the plasmon with other degrees of freedom are not well
understood.  One interesting question is the nonlinearities
associated with multiple plasmon excitations--how weakly do
they interact with each other?  Some physical processes 
can be sensitive to nonlinearities; 
for example ionization may be
energetically impossible for individual plasmons but allowed
for states with multiple plasmon excitations and therefore, ionization rates 
may depend on the degree of nonlinearity. 
Recently an
experiment was reported observing the ionization probability with 
different photon field durations\cite{sc98}. The photon energy was such
that ionization is energetically possible only if at least two
photons are absorbed.  In this work we ask whether the 
observed ionization can be interpreted as electron emission by a two-plasmon
state within a 
simple theory based on the jellium model
of the electronic structure.  We will use standard many-body perturbation
theory, describing the surface plasmon in RPA as a particle-hole excitation.
The plasmon description and details of the jellium model are given in
Sect.~2 below.

Our calculation may be viewed as a semianalytic approximation to 
time-dependent local density approximation (TDLDA), in which the 
electron dynamics is treated  entirely by a common potential field.
The TDLDA has been well developed for the high-field response of
atoms\cite{se87,to98,gr97}, and is now being applied to sodium 
clusters\cite{ul97}.  Unfortunately, the full TDLDA is computationally 
difficult and rather opaque, in contrast to the perturbative 
approach that allows important quantities to be calculated directly.

From the point of view of surface plasmon dynamics, a very important
quantity is the ionization rate of a two-plasmon
excited state.  
Haberland et al. \cite{sc98} interpreted their measurements under
the assumption that this rate is fast on a  time scale of 10 fs, and
we wish to see whether that can be justified theoretically.  The
two-plasmon ionization rate is calculated in Sect.~3.  However,  the
ionization can take place without the mediation of the plasmon.  Also, the
plasmon can be excited as a virtual state in which case the connection to 
the two-plasmon decay formula is unclear.  We present in Sect.~4 a more
general treatment of the ionization process that includes these effects
and allows the role of the plasmon to be isolated from other aspects of 
the ionization process. The important role of plasmon in screening and
enhancement of the external field is made explicit in the formulas
discussed there.

\section{The electronic structure model}
In this section we discuss the Hamiltonian model and the treatment of
the surface plasmon.  We will need single-electron
wave functions and energies, which we calculate as follows.  We first
obtained the solution of the self-consistent jellium model using the
computer code ``jellyrpa" \cite{be91}.  The jellium background charge
is assumed to be uniform in a sphere of radius $R=r_s N^{1/3}$.  Here
$r_s=3.93$ a.u. corresponds to density of charge equal to the bulk density
of atoms in sodium metal.  It is more convenient
for recreating the wave functions to work with analytic models of the
potential, so we fit the self-consistent jellium potential
to a Woods-Saxon shape.  
Specifically, we take
the electron potential to be 
\bea
V(r) = {-V_0 \over 1 + e^{(r-R_0)/a} } -V_c(r),
\label{eq1}
\eea
where $V_c(r)$ is a Coulomb field associated with the positive charge
distributed uniformly in the jellium sphere,
\bea
V_c(r) & = & {e^2\over r}~,\quad r>N^{1/3}r_s \nonumber\\
 & = & {e^2\over R}\Biggl({3\over 2} -{r^2\over 2R^2}\Biggr)~,\quad
 r<N^{1/3}r_s~. 
\label{eq2}
\eea
The parameters that fit this potential to the self-consistent
one are $V_0=5.71$~eV, $R_0=10.548$~\AA, and $a=0.635$~\AA.  The 
occupied energy levels of
this potential are within $0.2$ eV of the self-consistent
potential, which is certainly within the accuracy of the 
jellium model.
We find that the  cluster
has an ionization potential of 4.5 eV. Under the conditions of
the experiment [2] using photons of 3.1 eV, two photons are
required for ionization on energetic ground.  The
single-electron spectrum is shown in
Fig.~1. We use these orbitals and energies in the RPA and 
ionization calculations.

The RPA surface plasmon might also be calculated numerically
with the code ``jellyrpa", but in the interests of developing
analytic formulas we adopted a more schematic approach.  We
take the interaction between electrons to have a separable
form\cite{separ}, 
\bea
v({\bf r},{\bf r'}) = \kappa  {\bf f}({\bf r})\cdot {\bf f}({\bf r'})~,
\label{eq3}
\eea
where ${\bf f}$ is a three-dimensional vector with components
${\bf f}_\mu({\bf r})\equiv f(r)Y_1^{\mu}(\hat r)$. 
Then the energies of the RPA excitations satisfy the dispersion
relation
\bea
1 = 2\kappa \sum_{ph}{\langle p | {\bf f}_\mu| h \rangle^2 (\e_p - \e_h)
\over \omega^2-(\e_p - \e_h)^2}~,
\label{eq4}
\eea
where $\e$ is a single-particle energy and $p,h$ label 
particle and hole orbitals.  Due to the spherical symmetry, the 
solutions $\omega_n$ of the dispersion relation are independent of $\mu$.
The matrix element 
$\langle n\mu | {\bf f}_\mu | 0\rangle$
between the ground state and a one-phonon
state of energy $\omega_n$ is given by 
\bea
\langle n\mu | {\bf f}_\mu | 0\rangle & = & 
{{1}\over {2 \kappa}} \Biggl(\omega_n \sum_{ph} {{\langle p | {\bf f}_\mu 
    | h \rangle^2 
(\e_p - \e_h)} \over {(\omega_n^2-(\e_p - \e_h)^2)^2 }} \Biggr)^{-1/2}~.
\label{eq5}
\eea
We shall particularly require the transition potential $v_{n\mu}$ associated
with the creation of the plasmon.  This is given by
\bea
v_{n\mu} = \kappa {\bf f}_\mu({\bf r}) \langle n\mu | {\bf f}_\mu | 0\rangle~.
\label{eq6}  
\eea

In the spherical jellium model, the surface plasmon can be roughly 
described taking the interaction of dipole-dipole form.  For
an excitation along the $z$-axis, the field is
\bea
{\bf f}_0({\bf r})& = & z~.
\label{eq7}
\eea
Assuming that the transition density of the plasmon
is concentrated at the surface at radius $R$, the strength of
the interaction is obtained from the multipole expansion of 
the Coulomb interaction as
\bea
\kappa_c = {e^2\over R^3}~.
\label{eq8}
\eea
The dispersion relation can then be solved analytically\cite{be94} in the
limit $\omega>>(\e_p-\e_h)$ making use of the TRK sum rule.  The
result is the simple 
Mie surface plasmon formula,
\bea
\omega_n^2 = { e^2 \hbar^2 N\over m R^3}~.
\label{eq9}
\eea

The resulting energy is about 25\% higher than the empirical value
for sodium clusters, $\omega\approx 2.75$~ eV.  The RPA can be made to
fit this value for $N=93$ by renormalizing the coupling strength by
$\kappa=0.52\kappa_c$.
However, the transition potential of eq.(\ref{eq6}) calculated with
eq.(\ref{eq7}) is linear in $r$ whereas TDLDA calculations without separable
assumption do not yield this behavior, as shown in Fig.~2. A simple
improvement over the linear form eq.~(\ref{eq7}) is the dipole field associated
with a charge distribution localized on the surface of the jellium
sphere\cite{ya95}.  
A surface charge produces a radial field of the form
\bea
f(r) & = & r \quad r<R~, \nonumber\\
 & = & {R^3\over r^2}\quad r>R~.
\label{eq10} 
\eea
This is plotted as the dashed line in Fig.~2; it clearly is much closer
to the actual TDLDA transition potential. 
With this choice, the empirical position of the resonance is
obtained by using the coupling $\kappa=0.6\kappa_c$. This is very
close to the previous one, showing that the modified form factor
has only a small influence on the plasmon properties.  We will see
that it is much more important in the ionization process.    

In the experimental photoabsorption spectrum, the plasmon has a 
width of about 0.5 eV.  This finite width
requires theory beyond RPA, which produces only zero-width excitations
below the ionization threshold.  Since it is not easy to incorporate
the other degrees of freedom that are responsible for the width,
we will treat the width phenomenologically.  In discussing the response
in general, it is useful to consider the dynamic polarizability
$\alpha(\omega)$.  This is given by 
\bea
\alpha(\omega)  = \sum_n \frac {2e^2\langle n \vert z \vert 0 \rangle^2  
\omega_n}
{\omega_n^2 - (\omega - i \delta )^2}
\label{alpha}
\eea
where $n$ labels the true excitations and $\delta$ is a small quantity.
A simple prescription is to take a single pole for the plasmon, 
taking into account the finite width by the replacement $\delta\rightarrow
\Gamma_n/2$ for a width $\Gamma_n$.  

The imaginary part of $\alpha$ is related
directly to the photoabsorption cross section $\sigma$ by the formula
\be
\sigma = 4 \pi {\omega\over c} {\rm Im}~\alpha(\omega).
\label{sigma}
\ee
Given ${\rm Im}~\alpha(\omega)$, the real part can then be computed
from the Kramers-Kronig relation,
\be
{\rm Re} \alpha(\omega) = {1\over\pi} \int^\infty_0 d \omega'
{\rm Im} \alpha(\omega')\Bigg({2\omega'\over \omega^2-\omega'^2}\Bigg).
\label{KK}
\ee
Applying this to the experimental data of ref. \cite{re95}, we find
the imaginary part of $\alpha$ graphed as the solid line in Fig.~3.  
This is compared with the single-pole approximation with parameters
$\omega_n=2.75$ eV and $\Gamma_n=0.5$
eV.  A modification of the jellium model was 
proposed in ref.\cite{ca97} introducing a soft-edged surface in the distribution
of the background charge.  We also calculated the full RPA response
for soft jellium model, calculated using the program ``jellyrpa".  In
this case the empirical width can be reproduced with a smaller external
width parameter.  The dashed line shows the fit with $\Gamma_n=0.3$ eV.
Both these models give a reasonable but not quantitative description of
the data. The soft jellium  model has the advantage 
that the plasmon can be moved to lower frequency without adjusting 
the coupling strength.  However, it predicts too low an ionization
potential, which makes it unsuitable for the autoionization calculation.

The corresponding comparison for the real part of $\alpha$ is shown
in Fig.~(4).  Here the theory is quite robust, and we can rather 
confidently estimate Re $\alpha$ at the energy of interest (3.1 eV)
to be about 3000 \AA$^3$.

\section{Two-plasmon autoionization rate}

The many-body perturbative graphs for $M_{ph}$, the interaction matrix
element between the two-plasmon excitation of the $n\mu$ mode and the 
final configuration with a hole $h$ and the
electron in a continuum state $p$ , is shown in Fig.~5. 
The labels $p',h'$ stand for particle and hole
states, respectively.  Algebraically,
the matrix element is given by 
\bea
M_{ph} & = & \sqrt{2}\bigg[ 
\sum_{p'}
{\bigg(
  \frac 
  {\langle h |v_{n\mu}|p'\rangle\langle p' | v_{n\mu} | p\rangle } 
{\omega_n-\e_{p'}+\e_{h}}
\bigg)}
-\sum_{h'}{\bigg(\frac {\langle h |v_{n\mu}|h'\rangle
\langle h' | v_{n\mu} | p\rangle} 
{\omega_n-\e_p+\e_{h'}}\bigg)} 
\bigg] 
\nonumber\\
 & = &\sqrt{2}
 \sum_{i'}
 \bigg( \frac 
 {\langle h |v_{n\mu}|i'\rangle \langle i' | v_{n\mu} | p\rangle}
{\omega_n-\e_{i'}+\e_{h}}
\bigg)~,
\label{eq11}
\eea
where $v_{n\mu}$ is defined in eq.~(\ref{eq6}). 
The factor $\sqrt{2}$ accounts for the statistics of the
two-plasmon initial state.
The two graphs can be combined in one sum over both particles and
holes as shown in the second line, making use of
the fact that the matrix element is only required
on shell, i.e. with $\e_p - \e_h = 2 \omega_n$.
The primed indices $i'$ indicate
particle or hole orbitals, depending on the direction of the arrow.
The ionization width $\Gamma_e = \hbar w_e$, where $w_e$ is the ionization
rate, is given by the Golden Rule formula,
\bea
\Gamma_e & = & 2 \pi\sum_{ph} |M_{ph}|^2 {dn_p\over dE} \delta(2\omega_n -
\e_p + \e_h) \nonumber\\
 & = & 4 \pi \kappa^4 \langle \ n |{\bf f_{\mu}}| 0 \rangle^4 
 \sum_{ph}|K_{ph}|^2
{dn_p\over dE} \delta(2\omega_n -\e_p + \e_h)~.
\label{eq12}
\eea
Here $dn_p/dE$ is the density of states of the continuum electron.
We have also separated out the
excitation amplitude for a field $\bf f$,  
\bea
K_{ph} & = & \sum_{i'}\bigg({\langle h |{\bf f}_\mu|i'\rangle
\langle i'| {\bf f}_\mu | p\rangle\over 
\omega_n-\e_{i'}+\e_h}\bigg)~.
\label{eq13}
\eea

The sums in eq.~(\ref{eq12}) can be reduced in size by making use of the
angular momentum symmetry of the orbitals.  
Labeling the angular momentum quantum numbers $l$ and $m$, we
may express the $m$-dependence of the matrix elements as  
\begin{eqnarray}
\langle p',m_{p'}\vert {\bf f}_\mu \vert p, m_p \rangle & = &  
(-1)^{l_{p'}-m_{p'}}
 \left ( \matrix{ l_{p'} & 1 & l_{p} \nonumber \cr 
     -m_{p'} & \mu & m_{p} \nonumber
\cr } \right ) \langle p'\vert\vert  {\bf f} \vert\vert p\rangle~,\nonumber \\
\label{eq14}
\end{eqnarray}
where the reduced matrix element 
$\langle a \vert\vert  {\bf f} \vert\vert b \rangle$  is defined as\cite{bohrI}
\begin{eqnarray}
\langle a \vert\vert {\bf f} \vert\vert b \rangle & = &  (-1)^{l_a}
\sqrt{(2 l_a+1)(2 l_b +1)} \left( \matrix{ l_a & 1 & l_{b} 
  \cr 0 & 0 & 0 
\cr } \right ) \int_0^{\infty} f(r)\varphi_a(r) \varphi_b(r) r^2 dr~
\label{eq15}
\end{eqnarray}
in terms of the radial wave functions $\varphi_i$.  
The sum over magnetic quantum numbers $m_{p,h}$ implicit in eq.~(\ref{eq12}) 
can be evaluated in terms of a 9-j symbol\cite{ro59} in which the total
angular momentum $L$ carried by the two photons appears.  The
result is
\begin{eqnarray}
\sum_{all~m} |K_{ph}|^2 & = & 2\sum_{ij} \frac{1}
{\omega-(\epsilon_{j}-\epsilon_{h})}
 \frac{1}{\omega-(\epsilon_{i}-\epsilon_{h})}
 \langle h\vert\vert {\bf f} \vert\vert j \rangle 
\langle j\vert\vert {\bf f} \vert\vert p \rangle 
 \langle i\vert\vert {\bf f} \vert\vert h \rangle 
\langle p\vert\vert {\bf f} \vert\vert i \rangle \nonumber \\
 & \times & \sum_{L=0,2} {\hat L}^2 
\left ( \matrix{
 1 & 1 & L \nonumber \cr 0 & 0 & 0 \nonumber \cr }
\right )^2  
  \left\{ \matrix{ 
1 & 1 & L \nonumber \cr l_{j} & l_p & 1 \nonumber \cr l_{h} & l_{i} & 1 
\nonumber \cr } 
\right\}~.\\
\label{eq16}
\end{eqnarray} 
The factor of 2 arises from the two-fold spin degeneracy of the occupied
orbitals.  In carrying out the calculation one also has to fix the 
normalization of the continuum radial wave function. 
A convenient choice is
$r\varphi_p \rightarrow \sin(kr+\delta)$
at large $r$, giving $dn_p/dE = 2 m/(\pi k \hbar^2)$.

\def\na93{Na$^+_{93}$}
The results for the autoionization of \na93 are given in Table I.
We chose the plasmon parameter
$\kappa$ in two different ways, requiring the plasmon resonance energy
$\omega_n$
to be either at the experimental position of 2.75 eV, or at the
energy of the absorbed photons, 3.1 eV. The particle-hole states
taken into account in the calculations include
electron jumps up to three
harmonic oscillator shells. We first discuss the
results in the case of undamped excitations ($\delta=0$ in Table I).
The upper half of Table I shows
the values obtained using the linear dipole field ${\bf f}_0= z$
(eq.~(\ref{eq7})).
The resulting widths $\Gamma_e$ are extremely small for both choices of
$\omega_n$, and they would be hard to
reconcile with experiment.  This led us to reexamine our simplifying
assumption about the shape of the separable particle-hole interaction.
Since the choice of eq.~(\ref{eq10}) gives a better transition potential
(see fig.~2) we use it from now on instead of eq.~(\ref{eq7}). 
As shown in the lower half of Table I, the resulting 
widths are much larger and they seem to give a possibility to explain the 
data. Indeed, they correspond to ionization times of the order of 5.5 fs to 
7.5 fs which is comparable to the estimated plasmon lifetime of 10
fs\cite{sc98}.

However, the calculated results cannot be considered reliable because they
are quite sensitive to the single-particle energies involved
in the transition. In eq.~(\ref{eq13}) several states $i'$ give quite 
small energy denominators (see Fig.~1) and therefore they yield
abnormally large contributions. However, 
it is not consistent to neglect the damping of the excitations
in the perturbative  calculation when the energy denominators are small.  
As we did with the plasmon in Sect.~2, we here add a finite imaginary
term $\delta$ to the energy
denominators of eq.~(\ref{eq13}). In Fig.~6 we show the dependence of
$\sum |K_{ph}|^2$ on $\delta$ which is seen to be moderate for 
$\delta$ in the range
0.1 - 0.2 eV. 
We see that the damping
reduces the ionization width by roughly a factor of two when $\delta=0.1$ eV
which seems a reasonable value. 
Although the dependence on the transition field and on the damping
makes the calculation uncertain by a factor of two or so, we at least
see that the result is rather insensitive to the particular model for
the plasmon excitation energy.  This is nice in that it means that
the theoretical uncertainty here is not a hindrance to doing the
calculation.

The results correspond to a ionization lifetime $\tau_e$ of the 
two-plasmon state in the range of 5 to 10 fs. 
This is of the same order of magnitude as the plasmon lifetime which is
estimated to be about 10 fs\cite{sc98}. In the present model the ionization
process is not very fast contrarily to the assumption made in 
the interpretation of ref.~\cite{sc98}, but it is fast enough to allow for
the ionization process to compete with the plasmon damping. 

\section{Ionization in a laser field}
In this section we wish to apply the previous results to ionization in
a laser field.   Thus, we consider the ionization as a multistep process, in
which the
photons are first absorbed to make plasmons and then the plasmons
interact to eject an electron.
A simple physical argument
can be made to obtain a formula for the ionization, which we 
will then justify more formally.  Let us define the absorption rate 
for photons $R_\nu$
and the damping width for the plasmon $\Gamma_n$. In the steady
state  the balance 
between the creation and absorption of plasmons gives a mean number 
of plasmons $\bar n$ satisfying
\bea
\bar n & = & \frac {\hbar R_\nu}{ \Gamma_n}
\label{eq17}
\eea
Taking the distribution of numbers as Poissonian, the mean number of pairs
is then 
$\bar n^2/2$. The ionization rate $R_e$ 
is related
to the two-plasmon ionization width $\Gamma_e$ by
\bea
R_e & = & \frac {\bar n^2}{2} w_e \nonumber \\
    & = & \frac {\hbar}{2} \frac{R_{\nu}^2}{\Gamma_n^2}\Gamma_e~,
\label{eq18}
\eea
where $\Gamma_e$ and $w_e$ have been introduced in the preceding section.

A more formal derivation of this formula may be made from the graph of
Fig.~5
as follows.  We add to the graph matrix elements 
of the external field between the zero- and
one-plasmon states, 
\bea
V_n = \langle n | V_{ext}|0\rangle~.
\label{eq19}
\eea
The plasmon propagator itself can be approximated by
$1/( -\omega + \omega_n + i \Gamma_n/2)$ for $\omega$ close to $\omega_n$. 
Then the ionization rate for the graph is given by
\bea
R_e & = & \frac{2\pi}{\hbar} 
\sum_{ph}\Biggl| \Bigl({V_n\over  -\omega + \omega_n + 
  i \Gamma_n/2}
\Bigr)^2 M_{ph}\Biggr|^2{d n_p \over dE} \nonumber \\
 & = & \frac{1}{\hbar}
 {V_n^4\over ((\omega_n-\omega)^2 +(\Gamma_n/2)^2)^2}{\Gamma_e\over 2}~.
\label{eq20}
\eea
On the other hand, the photon absorption rate can be calculated as
the imaginary part of the self-energy associated with the
coupling $V_n$,
\bea
R_\nu & = & \frac{-2}{\hbar} V_n^2 {\rm Im}{ 1 \over -\omega + \omega_n + i
  \Gamma_n/2}\nonumber \\
 & = & \frac{1}{\hbar} 
 {V_n^2 \Gamma_n\over (\omega_n-\omega)^2+(\Gamma_n/2)^2}~.
\label{eq21}
\eea
Eq.~(\ref{eq18}) may now be obtained by combining the last two equations.  

For comparing with experiment, it is convenient to express eq.~(\ref{eq18})
in terms of the
number of ionizations per cluster $N_e=R_e T$ and the number of
photons absorbed $N_\nu=R_\nu T$, where $T$ is the time duration of the
laser pulse. We obtain for $N_e$
\bea
P_e & \equiv & \frac {N_e}{N_{\nu}^2/2} \nonumber \\
 & = & \frac {\hbar}{T} \frac {\Gamma_e}{\Gamma_n^2}~.
\label{eq22}
\eea

The experiment [2] observed not only ionization but considerable
evaporation of atoms from the clusters.  We note that atomic
motion takes place on a much longer time scale than electronic
motion.  Most of the evaporated particles are emitted from the
cluster after it has reached thermal equilibrium; the statistical
evaporation theory gives lifetimes in the nanosecond regime for
the conditions of the experiment\cite{evap}.  This is longer
than the fast laser pulse duration ($T=140$ fs),  so
we may think 
that the \na93 cluster remains whole before being ionized.
The first line of eq.~(25) may be estimated from Fig.~2 of 
ref. \cite{sc98}.  The broad peak in the middle of the mass spectrum
represents ionized clusters and has roughly 1/3 of the area of the
\na93 peak on the right; thus $N_e\approx1/3$. The
number of photons absorbed is given as $N_\nu\approx 6$.  Thus we
estimate\footnote{Fitting more extensive data with a rate equation,
the authors of ref.\cite{sc98} find a somewhat larger probability,
$P_e\approx 0.06$} $P_e\approx 0.014$.  

To evaluate the second line of eq.~(25) we have the calculated
$\Gamma_e$ from Sect.~3 and the experimental pulse duration $T$, but
we still do not have $\Gamma_n$, the damping width of the plasmon.
As discussed in Sect.~2, this quantity is beyond the scope of RPA,
but an upper bound to
$\Gamma_n$ is given by the empirical photoabsorption spectrum.  Taking
the values $\omega_n$=2.75 eV and $\Gamma_n=0.5$ eV the single-pole fit
to the plasmon and $\delta=0.2$ eV  in the perturbative energy
denominators, we find $P_e= 1.0\times 10^{-3}$, which is much too small.
On the other hand, the experimentally quoted lifetime of the plasmon is 10
fs which corresponds to a considerably smaller value of $\Gamma_n$. Adopting
this value of $\Gamma_n$ gives $P_e= 5.8\times 10^{-2}$. The 
width $\Gamma_e$ depends on the particular cluster considered since the
energetics of the emitted electrons will change somewhat for different
clusters. 

\section{General theory}
The theory in the last section assumed that the energy transfer to
the electron was indirect, first producing plasmons which then autoionize.
In fact
the photon could be absorbed directly on the
electron.  The direct absorption is implicit in the TDLDA, and 
can be taken into account as well in the perturbative theory
of photon absorption\cite{be66}.
The general expression for a second-order transition
from a state $i$ to $f$ is given by
\bea
w_{fi} & = & \frac {2\pi}{\hbar} e^4 {\cal E}_0^4 |K_{fi}|^2 {d n_f\over d
  E}~.
\label{eq23}
\eea
The second order matrix element $K_{fi}$ is similar to eq.~(16) with
the linear field (7), except the particle orbitals are replaced by
many-body states $i,i',f$: 
\bea
K_{fi} & = & 
\sum_{i'}{\langle f | z|i'\rangle
\langle i' | z| i\rangle\over 
\omega-E_{i'}+E_i}
\label{eq24}
\eea
In eq.~(\ref{eq23}) $ {\cal E}_0 $ is the amplitude of the time-dependent 
electric dipole 
field, $\vec{\cal E}(t)={\cal E}_0 {\bf z} (e^{-i\omega t}+e^{i\omega t})$. It
is related to
the laser intensity by ${\cal E}_0^2= 2\pi I \omega/c$.
Taking the many-body states as the simple particle-hole configurations,
eq.~(\ref{eq24}) reduces to eq.~(\ref{eq13}). 
By itself this would only give the ionization probability of the electron in
the external field, i.e. without plasmon effects.  The many-body physics is
included by considering higher-order perturbations in the wave functions
including excitation of the other electrons by the active electron.  
The
result is to replace the external field $ {\cal E}_0 $ in eq.~(\ref{eq23})
by an effective field such that
\bea
 e{\cal E}_{eff}z & = & e{\cal E}_0 z - e \int v(r,r') \Pi(r',r") 
 {\cal E}_0 z" d^3r'd^3r"~,
\label{eq25}
\eea
where $\Pi$ is the response function of the cluster.  

To make the
connection with the previous approach, we approximate the response by a
single pole and use the separable approximation.  The effective field becomes
\bea
e{\cal E}_{eff}z & = & e{\cal E}_0 z - e {\cal E}_0 \kappa z  
\langle n |z|0\rangle\langle 0 |z|n \rangle\Bigl({1\over
-\omega + \omega_n + i \Gamma_n/2}+{1\over
\omega + \omega_n - i \Gamma_n/2}\Bigr) \nonumber \\
 & = & e{\cal E}_0 z \Bigl(1 - \frac {\kappa}{e^2} \alpha(\omega) \Bigr)~.
\label{eq26}
\eea
Eq.~(\ref{eq20}) can now be obtained from eq.~(\ref{eq26}) by
dropping the external field contribution as well as the nonresonant
term in the polarizability. 
In the second line, we express the polarization effects as a 
multiplicative factor, which can be interpreted as an effective
charge
\bea
 e_{eff} & = & e (1 - {\kappa\over e^2} \alpha(\omega))~.
\label{eq27}
\eea
We may use this expression to assess the relative importance of the
external and induced fields.  
First note that there is a complete cancelation of the two terms in 
eq.~(30), implying complete screening, if we take
$\kappa=\kappa_c$ and $\alpha=R^3$.  This last is just the classical
polarizability of a conducting sphere, and it may also be derived putting
the Mie resonance eq.~(\ref{eq9}) into the polarizability formula
eq.~(\ref{alpha}).
 
For the present purposes we assume in eq.~(29)  $\omega_n=2.75$ eV and 
$\kappa=0.52\kappa_c$ as
in Sec.~2. We also take  
for $\Gamma_n$ a width 
corresponding to a lifetime of 10 fs.  The result at $\omega = 3.1$ eV is
\bea
{\kappa\over e^2}\alpha(3.1~{\rm eV}) = - 3.3 +i0.3
\label{eq29}
\eea
Thus the plasmon enhances the field approximately by a factor of 4.3,
showing that the induced field is indeed dominant. 
This result is insensitive to $\Gamma_n$ if it is small,
but would
decrease if the width were as large as the measured 
width of the optical absorption
peak.  In a study of 
Na$_9^+$ in the TDLDA, ref. \cite{re98} obtained enhancement  factors 
on resonance in the
range 5-8, which is the same order of magnitude.  

The effect on the ionization rate goes as the fourth power of the field,
\bea
\vert 1 - \frac {\kappa \alpha(3.1~{\rm eV)}}{e^2}\vert^4 & = & 320~,
\label{eq30}
\eea
Finally , we use this result to make an
improved calculation of the ionization probability
$P_e$ introduced in eq.~(\ref{eq22}). The number of emitted
electrons is calculated as $N_e = w_{fi}T$ where $w_{fi}$ is given by
eq.~(\ref{eq23}) with $e$ replaced by $e_{eff}$.  
The number of of pairs
of absorbed photons is given by
$N_{pair}=(I\sigma T)^2/2$, $\sigma$ being the
photo-absorption cross section.  Computing $\sigma$ from eq.~(\ref{sigma})
we find $P_e = 1.25 \times 10^{-2}$, a
magnitude comparable to that found in the preceding section. However,
the cross section corresponding to the polarizability in eq.~(\ref{eq29})
is much too small at 3.1 eV, as is clear from the width discussion in
Sect.~2. On the other hand, if we take the width parameter from
empirical single-pole fit, $\Gamma_n$ = 0.5 eV, the 
probability comes out very small, as discussed earlier.    

\section{Conclusion}
In this work, we have derived the theory of cluster ionization 
by multiple photons of frequency near that of the surface plasmon.
The weak coupling between the surface plasmons is the driving 
interaction for the two-photon ionization process,
and a perturbative framework with respect to the ionized electron
seems reasonable.  The plasmon-induced mechanism can be derived from
the general perturbative formula using the higher order contributions
associated with the screened interaction.  Unfortunately, the formula
depends quadratically on the damping rate of the plasmon, which is
still not fully understood.   The rates obtained for \na93 are of the order of 
tens of femtoseconds, which is the same time scale as other relaxation
processes.  

We have used the jellium model in the calculations, and it is unclear how
realistic the model is.  We found that the interaction must be treated more
accurately than in the small-$r$ separable approximation, but we have not
examined the most sophisticated treatment of the interaction which would
include the exchange interaction without approximation. The Landau damping
of the Mie resonance is much larger with more realistic
interactions\cite{ma95}. A major problem of the jellium model is that the
damping is too small. Ionic scattering (called "interband transitions" in
condensed matter physics) would increase the spreading width, and the lack
of ionic scattering in the jellium model is in general a serious deficiency.
However, in the case of Na clusters the realistic Hamiltonian gives a very
similar spectrum to the jellium model\cite{ya96}, giving some credibility to
the model.  Nevertheless, it would be interesting to see what the effects of
the ionic scattering are in the second-order ionization. For small Na
clusters, more realistic calculations are becoming available of high field
ionization using the TDLDA \cite{re98b}, and it would be interesting to
compare.

\section{Acknowledgment}
We thank P.G. Reinhard for discussions, and G.B. thanks the IPN at Orsay for its hospitality where this work
was done.  He is also supported in part by the US Department of 
Energy under Grant DE-FG-06-90ER40561.


\begin{table}
\caption{Two-plasmon ionization widths in Na$_{93}^+$. The upper half and
  lower half of the table correspond to results calculated with the coupling
  (\ref{eq7}) and (\ref{eq10}), respectively. The quantities $\tau_e$ 
  are defined in the text.}
\begin{tabular}{c|ccc|ccc}
   & & $\omega_n$ = $2.75$ eV & & & $\omega_n$ = $3.1$ eV &  \\
   & & $\kappa$ = $0.91\times 10^{-2}$ eV$\AA^{-2}$ & & & $\kappa$ =
   $1.19\times 10^{-2}$ eV$\AA^{-2}$ & \\ 
\tableline
 $\delta$(eV) & $0.$ & $0.1$ & $0.2$ & $0.$ & $0.1$ & $0.2$ \\
 $\Gamma_e$(eV) & $9\times 10^{-4}$ & $5\times 10^{-4}$ & $4\times 10^{-4}$
 & $8.6\times 10^{-3}$ & $2.8\times 10^{-3}$ & $1.7\times 10^{-3}$ \\
\tableline
\tableline
   & & $\omega_n$ = $2.75$ eV & & & $\omega_n$ = $3.1$ eV &  \\
   & & $\kappa$ = $1.03\times 10^{-2}$ eV$\AA^{-2}$ & & & $\kappa$ =
   $1.34\times 10^{-2}$ eV$\AA^{-2}$ & \\ 
\tableline
 $\delta$(eV) & $0.$ & $0.1$ & $0.2$ & $0.$ & $0.1$ & $0.2$ \\
 $\Gamma_e$(eV) & $8.8\times 10^{-2}$ & $5.5\times 10^{-2}$ &
 $3.8 \times 10^{-2}$
 & $12.\times 10^{-2}$ & $5.8\times 10^{-2}$ & $5.\times 10^{-2}$ \\
$\tau_e$ (fs) & $7.5$ & $12.$ & $17.4$ & $5.5$ & $11.4$ & $13.2$ \\
\end{tabular}
\end{table}

\begin{figure}
  \begin{center}
    \leavevmode
    \parbox{0.9\textwidth}
      {\psfig{file=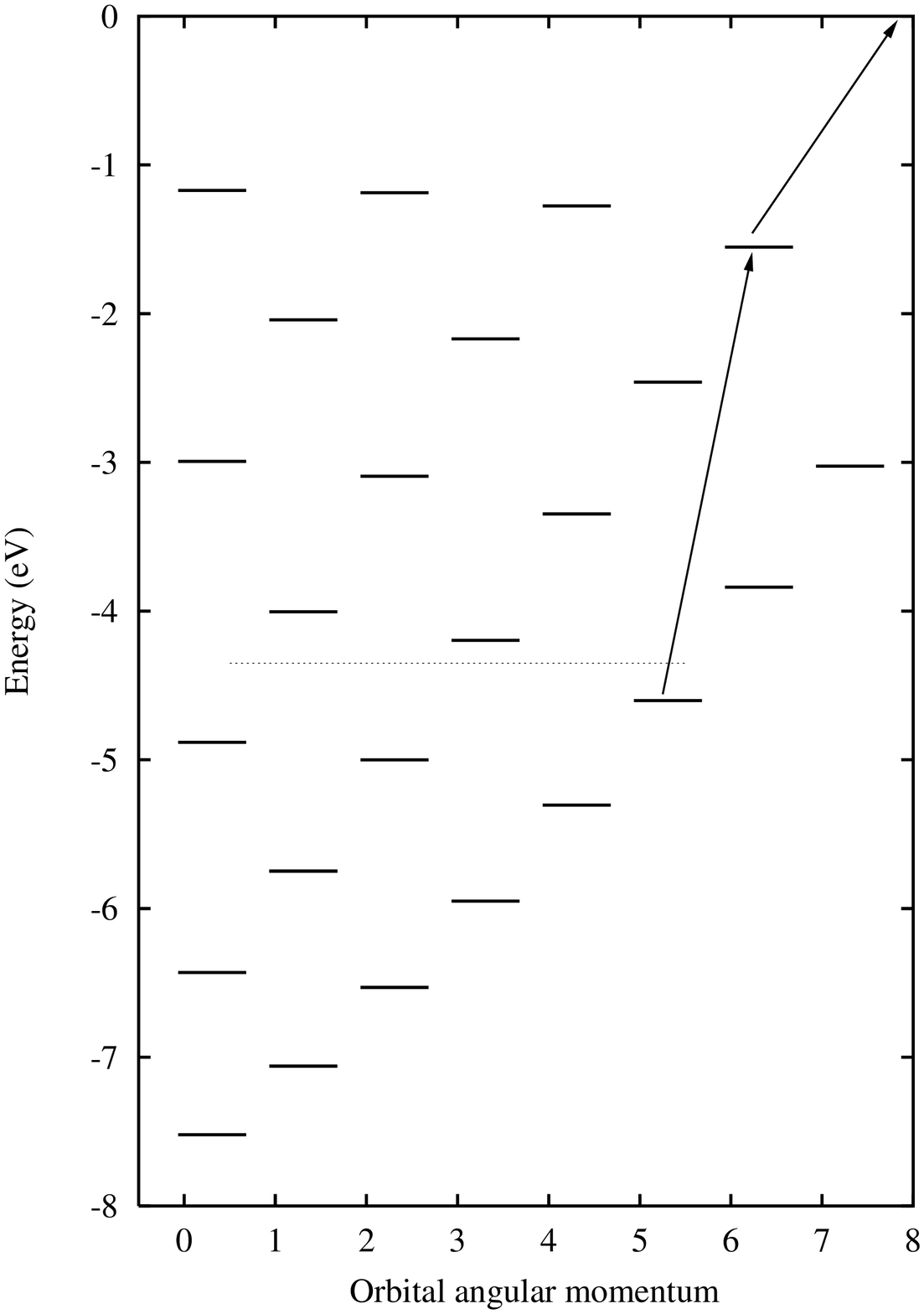,width=0.9\textwidth}}
    \end{center}
\caption{Single-particle levels in the jellium model of Na$_{93}^+$.
The Fermi level is indicated with a dotted line. The arrows show a
two-step transition with a particularly small matrix element in eq.~(16).}
\end{figure}

\begin{figure}
  \begin{center}
    \leavevmode
    \parbox{0.9\textwidth}
      {\psfig{file=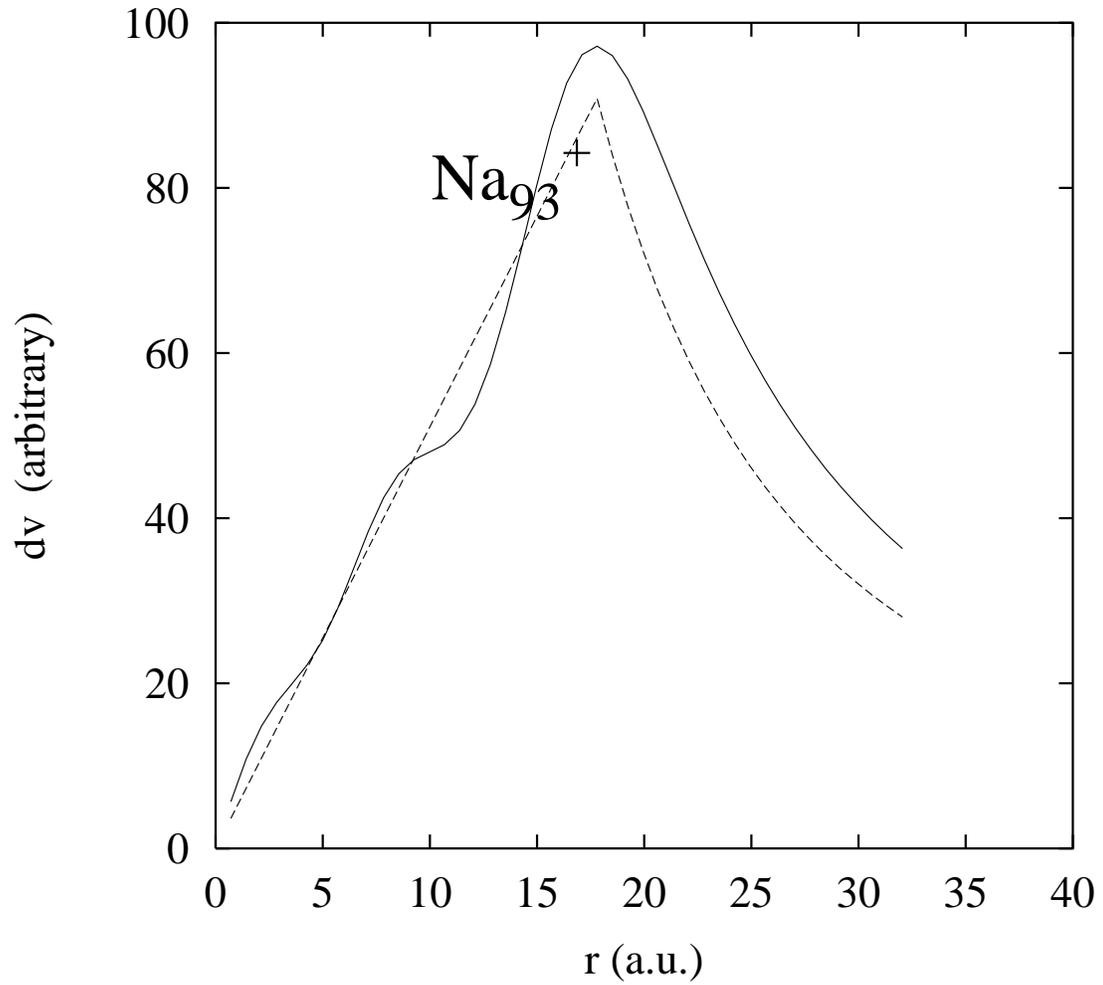,width=0.9\textwidth}}
    \end{center}
\caption{Transition potential, comparing full RPA with
eqs.~(\protect\ref{eq6},\protect\ref{eq10})}
\end{figure}

\begin{figure}
  \begin{center}
    \leavevmode
    \parbox{0.9\textwidth}
      {\psfig{file=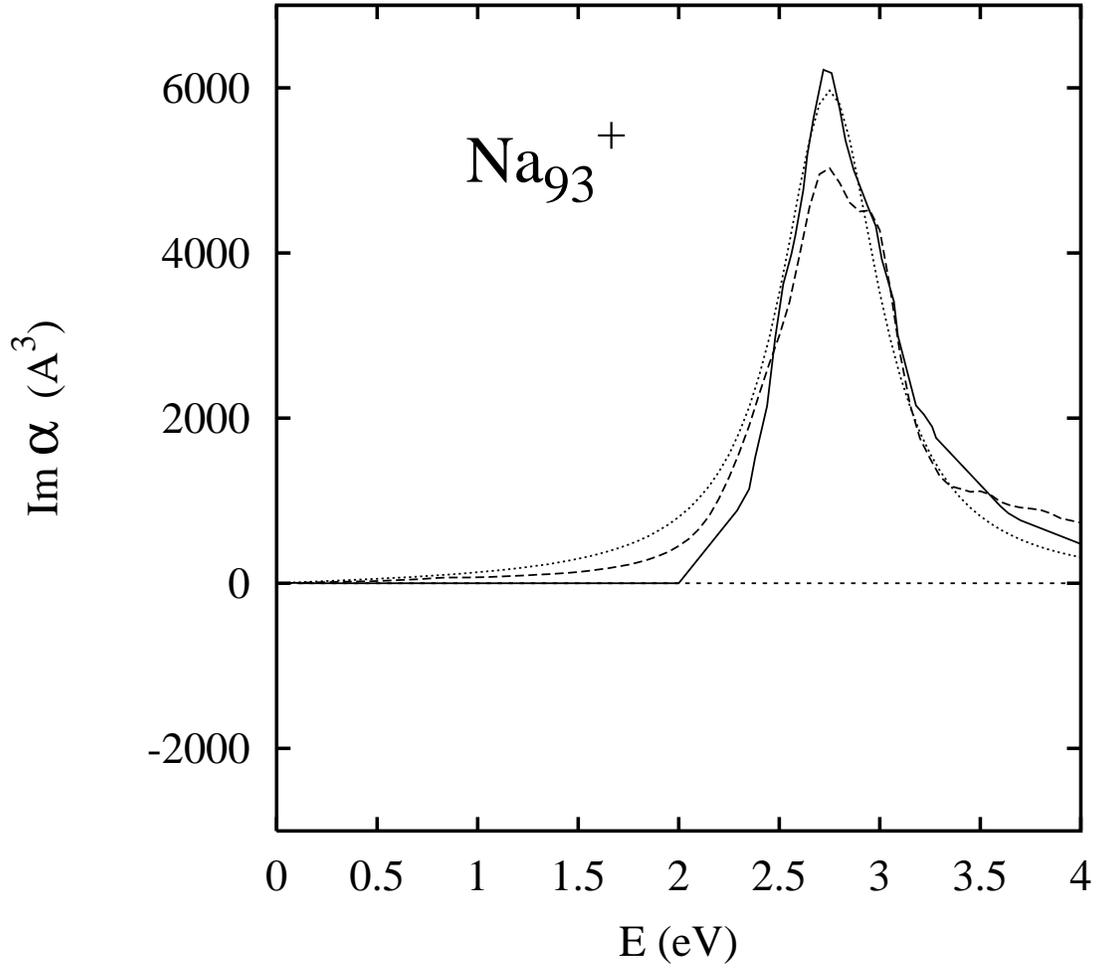,width=0.9\textwidth}}
    \end{center}
\caption{Imaginary part of the dynamic polarizability of Na$_{93}^+$:
empirical from ref. \protect\cite{sc98} and eq.~(\protect\ref{sigma}) (solid line); single-pole 
approximation (dashed line); RPA with soft jellium model (dotted line).}
\end{figure}

\begin{figure}
  \begin{center}
    \leavevmode
    \parbox{0.9\textwidth}
      {\psfig{file=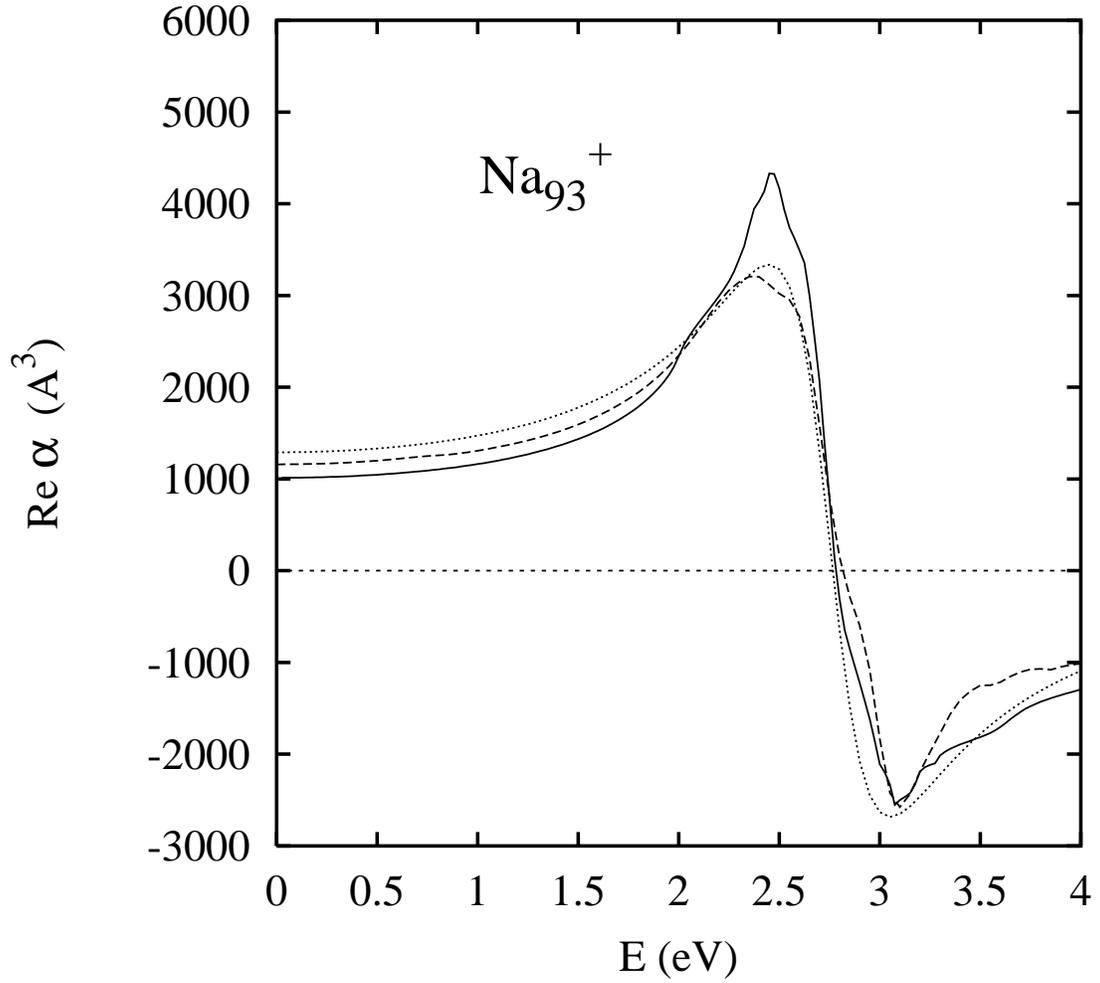,width=0.9\textwidth}}
    \end{center}
\caption{Real part of the dynamic polarizability for Na$_{93}^+$.
Empirical from Kramers-Kronig relation eq. (\protect\ref{KK}), solid line; 
single-pole approximation (dashed line); RPA with soft jellium
model (dotted line).}
\end{figure}

\begin{figure}
  \begin{center}
    \leavevmode
    \parbox{0.9\textwidth}
      {\psfig{file=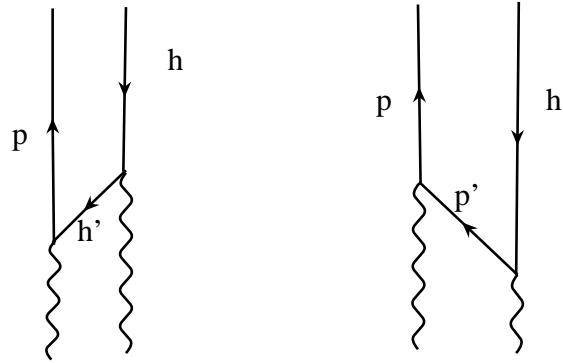,width=0.9\textwidth}}
    \end{center}
\caption{Perturbation theory graphs for second-order ionization.}
\end{figure}

\begin{figure}
  \begin{center}
    \leavevmode
    \parbox{0.9\textwidth}
      {\psfig{file=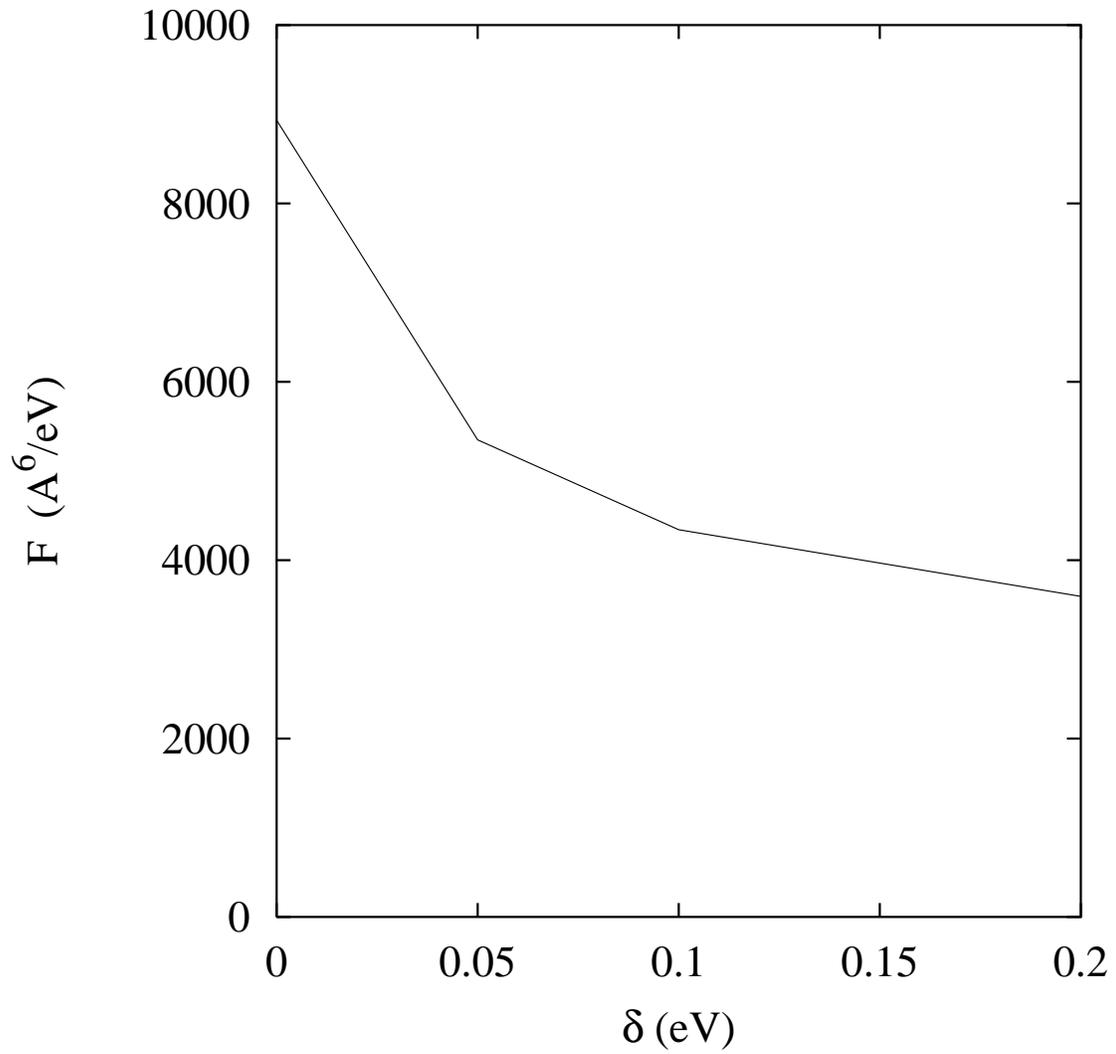,width=0.9\textwidth}}
    \end{center}
\caption{Second-order term $\pi e^4/2|K^{(2)}|^2dn_f/dE$
as a function of $\delta$.}
\end{figure}

\end{document}